\documentclass{aa}
\usepackage{txfonts}
\usepackage{graphicx}

\newcommand*{\flux}[1]{F_{\!{#1}}}

\newcommand*{\epstar}{\epsilon_*}
\newcommand*{\phistar}{\Phi_*}
\newcommand*{\eps}{\epsilon}

\begin{document}
%

\title{The spectral evolution of impulsive solar X-ray flares.
II.~Comparison of observations with models}

\titlerunning{Spectral evolution of solar X-ray flares. II.}

\author{P. C. Grigis \and A. O. Benz}

\offprints{\\Paolo C. Grigis,
\email{pgrigis@astro.phys.ethz.ch}}

\institute{Institute of Astronomy, ETH Z\"urich, 8092 Z\"urich, Switzerland}

\date{Received  27 October 2004 / Accepted 09 January 2005}

\abstract{
We study the evolution of the spectral index and the normalization (flux) of
the non-thermal component of the electron spectra observed by RHESSI during 24
solar hard X-ray flares. The quantitative evolution is confronted with the
predictions of simple electron acceleration models featuring the
soft-hard-soft behaviour. The comparison is general
in scope and can be applied to different acceleration models, provided
that they make predictions for the behavior of the spectral index as a
function of the normalization. A simple stochastic acceleration model
yields plausible best-fit model parameters for about 77\% of the 141 events
consisting of rise and decay phases of individual hard X-ray peaks.
However, it implies unphysically high electron acceleration
rates and total energies for the others. Other simple acceleration models such
as constant rate of accelerated electrons or constant input power have a
similar failure rate. The peaks inconsistent with the { simple} acceleration
models have smaller variations in the spectral index.
The cases compatible with a simple stochastic model require typically a
few times $10^{36}$ electrons accelerated per second at a threshold energy
of 18 keV in the rise phases and 24 keV in the decay phases of the flare peaks.

}

\maketitle

%
\section{Introduction}
%

The intense hard X-ray emission observed during solar flares is the
direct signature of the presence of highly energetic supra-thermal electrons.
The quest for a viable particle acceleration mechanism
drives both theoretical and observational investigations. In his review on
particle acceleration in impulsive solar flares Miller (\cite{miller98})
presents ``the major observationally-derived requirements'' for particle
acceleration.
The two main observational facts cited by Miller and used to put
constraints on the electron acceleration mechanisms are the time scales
of the acceleration (1~s for acceleration from thermal energies to 100~keV)
and the acceleration rates ($10^{36}$ to $10^{37}$ electrons
s~$^{-1}$ accelerated above 20~keV, to be sustained for several tens
of seconds).
Interestingly, the shape of the accelerated electron spectrum and its evolution
in time are barely mentioned. Grigis \& Benz
(\cite{grigis04}, henceforth Paper I)
have analyzed X-ray observations looking for systematic trends in the spectral
evolution of 24 impulsive solar flares { and confirmed the} predominant
\emph{soft-hard-soft} (SHS) behavior of the observed photon spectra,
{ first noted by Parks \& Winckler (\cite{parks69})}. This not
only applies to the global evolution, but is even more pronounced in individual
peaks. The authors also give a simple quantitative { description of the
SHS pattern, deriving an empirical} relation between the
normalization of the non-thermal component of the photon spectrum and its
spectral index. Can this systematic trend in the evolution of the hard X-ray
photon spectrum be used to put constraints on acceleration mechanisms?
As pointed out already in Paper I, the SHS behavior contradicts the idea that
the flux evolves by a varying rate of identical, unresolved events, termed
`statistical flare' in avalanche models (Lu \& Hamilton \cite{lu91}).

Here we study the constraints the new quantitative information on the SHS
behavior puts also on other acceleration models. Some conventional simple
acceleration scenarios (presented in Section \ref{acc_scenario}) are tested
whether they can reproduce the observed spectral behavior and what
constraints on their parameters can be obtained.
Our goal is to demonstrate the method, to stimulate further comparisons
between observation and theory and to call attention to the fact that the
spectral evolution cannot be neglected by a successful acceleration theory.
While the scenarios presented here are admittedly simple, we think that this
first step will be extended in the near future to encompass more sophisticated
models.

The main piece of information that we use for this comparison
is the relation between the normalization of the non-thermal component
of the spectrum (assumed to be a power-law) and its spectral index. 
Paper I studied this relation for photon spectra.
Here, we go a step further, and recover the electron spectra assuming
an emission from a thick target, which also yields a power-law spectrum
for the electrons.
Therefore we can use the data from Paper I and convert the photon
spectral index and normalization to the corresponding values for the
electron spectra.  This yields discrete time series of the spectral index
$\delta(t)$ and the power-law normalization $\Phi_{\epsilon_0}(t)$ at
energy $\epsilon_0$.
The acceleration models described in Section \ref{acc_scenario} provide
theoretical functions $\Phi_{\epsilon_0}(\delta)$ depending on the
model parameters, which can be fitted to the observed pairs
$(\delta(t),\Phi_{\epsilon_0}(t))$.
Applying this repeatedly for different flares and different emission
peaks during flares, the distributions of the best-fit model
parameters can be derived.

We summarize in Section \ref{data} the data reduction process yielding the
dataset. It is used in Section \ref{comparison} for the comparison with the
models described in Section \ref{acc_scenario}.

%
%

\begin{figure*}
\centering
\includegraphics{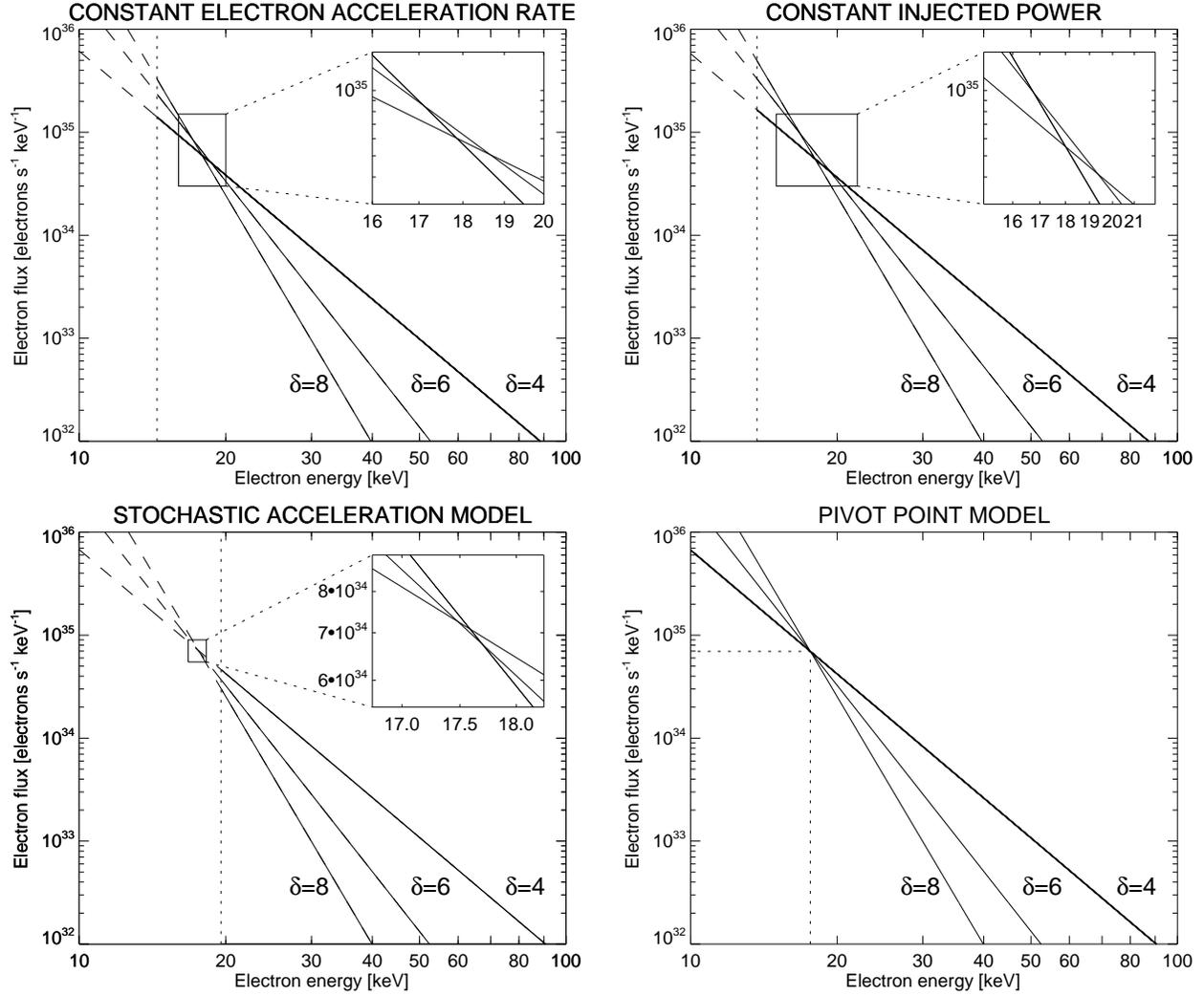}
\caption{Electron energy spectra with spectral indices of
$\delta =$ 4, 6, 8 for each of the four models indicated on top and
presented in Section \ref{acc_scenario}.
The vertical dashed line represents the lower energy threshold $\epstar$ for
the first 3 models, and the pivot point energy in the last plot.}
\label{model_spectra}
\end{figure*}
\begin{figure*}
\centering
\includegraphics{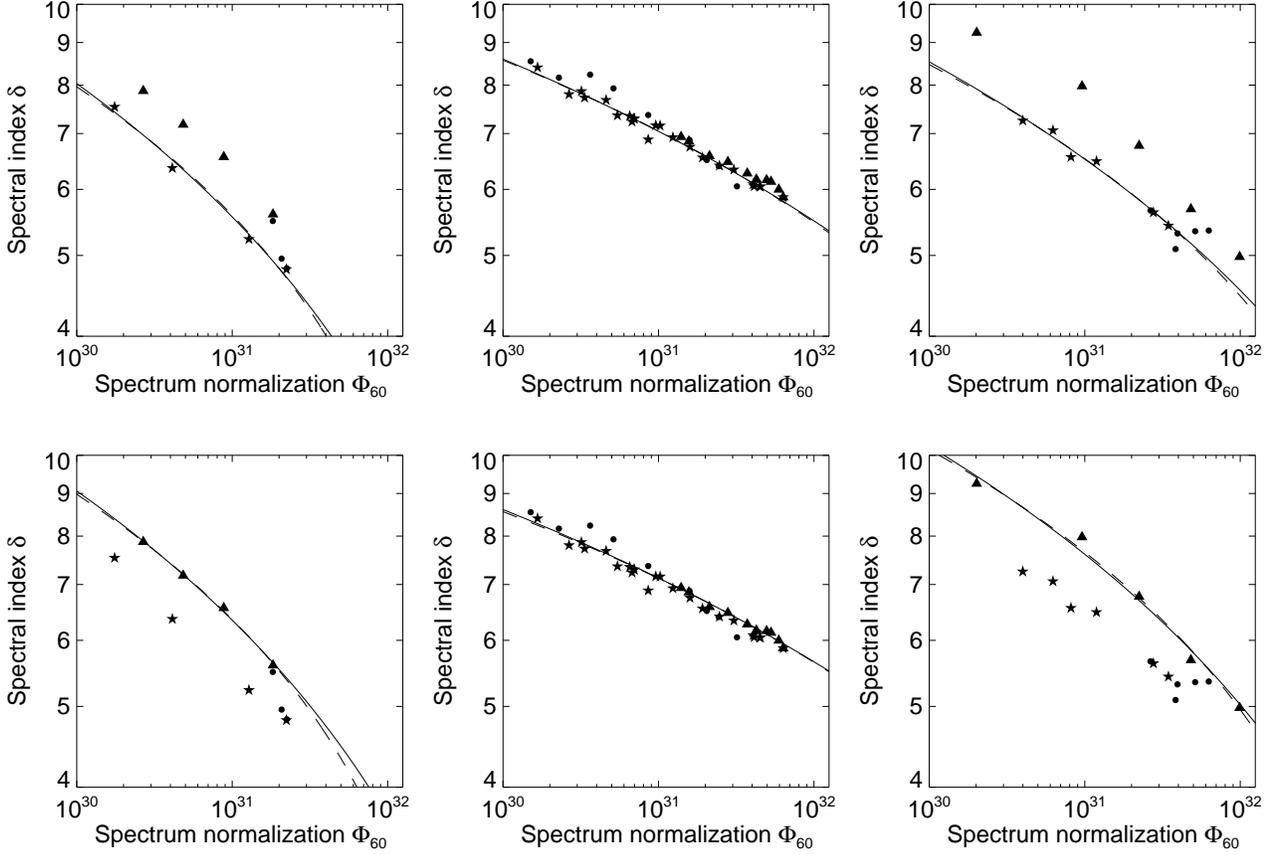}
\caption{Plot of the spectral index $\delta$  vs. the spectrum normalization
$\Phi_{60}$ for different times in three flares (first column: 20 Feb. 2002
09:54, second column: 04 Apr. 2002 15:29, third column 17 Apr. 2002 00:39).
The points belonging to a certain rise phase are represented by stars,
the points belonging to a decay phase are represented by triangles and the
other points are marked by dots.
The evolution of the constant rate and the stochastic acceleration model
are represented by the dashed and full lines, respectively. In the upper row
they were fitted to the rise phase points, in the lower row to the decay
phase points.}
\label{datamodel}
\end{figure*}
\begin{figure*}
\centering
\includegraphics{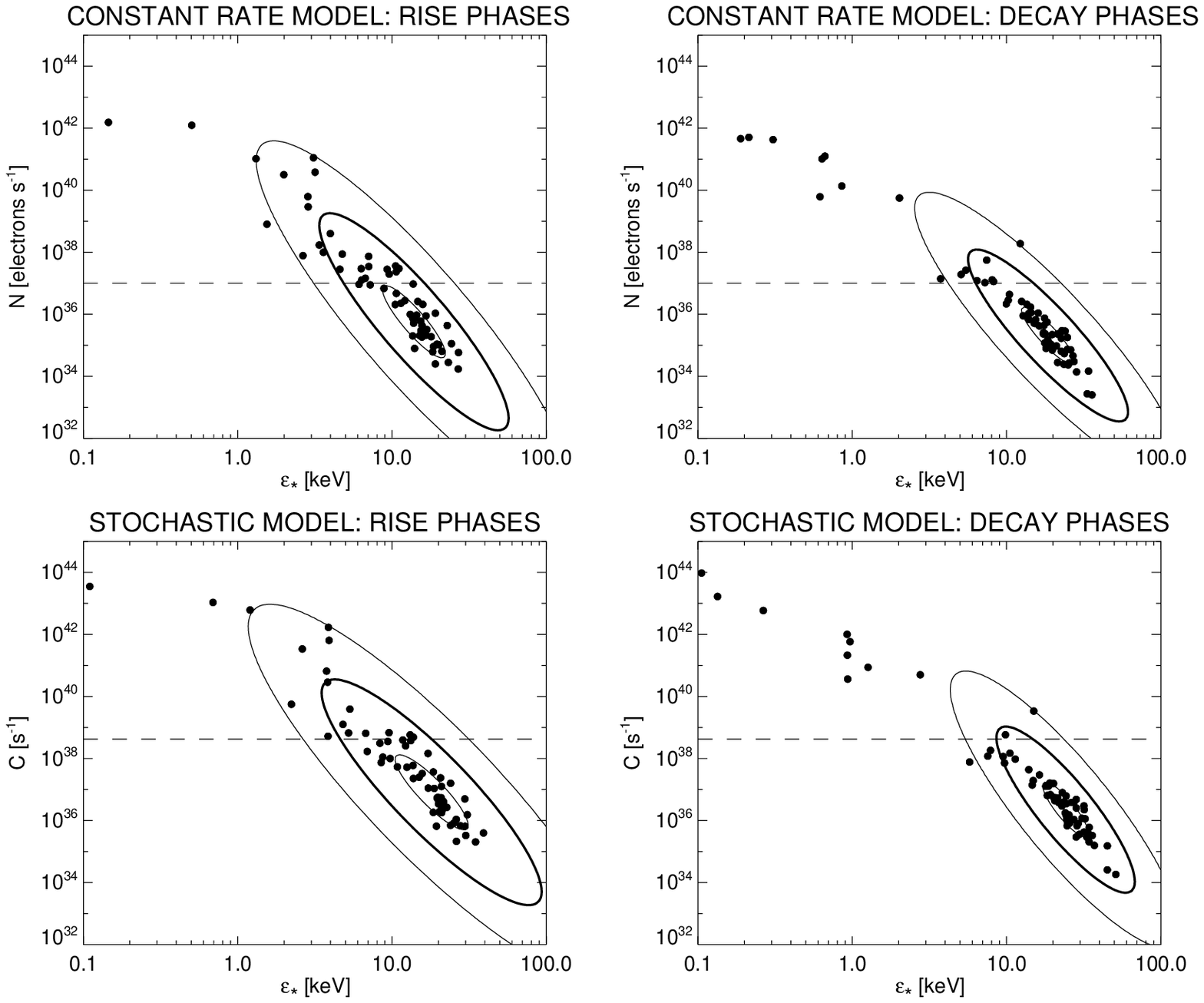}
\caption{Best-fit model parameters for the stochastic
acceleration and the constant rate model, separately for the rise and decay
phases. Contour lines corresponding to $1\sigma$,$3\sigma$ (thick line) and
$5\sigma$ levels for a Gaussian peak fitted to the data are also shown.
The region under the dashed line has an electron acceleration rate lower
than $10^{37}$ electrons s$^{-1}$ for events with spectral index $\delta = 6$.}
\label{twodimdist}
\end{figure*}


%
\section{Observations and Data Reduction}\label{data}
%

We give here a brief summary of the data reduction process, described
in full detail in Paper I. The photon spectral data used for this work is
exactly the same as the dataset used in Paper I, where hard X-ray observations
from the Reuven Ramaty High Energy Solar Spectroscopic Imager (RHESSI) of 24
solar flares of GOES class between M1 and X1 have been analyzed.
For each event count spectra were generated with a cadence
of one RHESSI rotation (amounting to about 4 seconds).
Each spectrum was fitted with a model consisting of an isothermal emission
(bremsstrahlung continuum and atomic emission lines) characterized by a
temperature $T$ and an emission measure $\mathcal{M}$, and a non-thermal
component characterized by a power-law with index $-\gamma$, normalization
$\flux{E_0}$ at energy $E_0$, and low-energy turnover at energy
$E_\mathrm{turn}$. The full detector response matrix was used to calibrate
the spectra.
The fittings were done by means of an automatic routine,
but were checked one by one, { excluding cases where the thermal and
non-thermal emissions could not be reliably separated, or the non-thermal
part was not well represented by a single power-law.}
After this selection we had a total of 911 good fittings for 24 events.
For this work, we just use the time series of the spectral index $\gamma$
and the normalization of the power law $\flux{E_0}$ for all the 24 events.
The non-thermal component of the spectrum is thus approximated by
\begin{equation}
  F(E)=\flux{E_0}\left(\frac{E}{E_0}\right)^{-\gamma}\, ,
\end{equation}
where $F(E)$ is the photon flux at 1~AU in photons s~$^{-1}$ cm~$^{-2}$
keV~$^{-1}$.
We now go one step further than Paper I and transform the photon
spectra into electron spectra. We choose the well-known 
analytically solvable thick target impact model (using the non-relativistic
Bethe-Heitler cross section and collisional energy losses) to
recover the injected electron spectrum. It is still a power-law
\begin{equation}\label{powerlawel}
\Phi(\epsilon) = \Phi_{\epsilon_0}
\left(\frac{\epsilon}{\epsilon_0}\right)^{-\delta}\,,
\end{equation}
where $\Phi(\epsilon)$ is the total number of electrons s~$^{-1}$ keV~$^{-1}$
over the whole target. The electron spectral index is given by
\begin{equation}
\delta =\gamma + 1
\end{equation}
and the electron spectrum normalization is
\begin{equation}
\label{equivalent_electron_spectrum}
\Phi_{\epsilon_0}= K\,\flux{E_0}\,E_0^\gamma\,\epsilon_0^{-\delta}\,
\frac{(\delta-1)\,(\delta-2)}{\beta(\delta-2,1/2)},
\end{equation}
where $\beta(x,y)$ is the beta function, and the constant $K$ is given by
\begin{equation}
K=\frac{3\pi^2 e^4 \ln \Lambda\,D^2}{Z^2\,\alpha r_e^2\,m_e c^2}
\simeq 6.4 \cdot 10^{33}
\,\mathrm{keV}\,\mathrm{cm}^2
\end{equation}
(Brown \cite{brown71}).
This transformation yields the time series $\Phi_{\epsilon_0}(t)$ and
$\delta (t)$, which will be used for the comparison with the acceleration
models.

The transformation of the observed photon spectrum into an electron spectrum
can potentially alter the results of the comparison between observations and
models, since it requires a knowledge of the physical conditions in the
emission region{, of their evolution in time} and of the importance of the
energy loss processes.
However, this is a necessary step for the comparison with acceleration models.
{ For simplicity we have neglected the effects of photon reflection in
the photosphere (albedo) and nonuniform target ionization.

The assumption of a thick target leaves open the origin of temporal variations.
In the following, we will assume that they are due to the acceleration rather
than to a variable release of trapped particle.}

%
\section{Acceleration scenarios}\label{acc_scenario}
%

We present here the frameworks of two acceleration scenarios, the
\emph{constant productivity scenario} and the
\emph{stochastic acceleration scenario},
which we will compare with the data on spectral flare evolution.
A scenario can comprise different models. All models are required
to yield a power-law distribution $\Phi$ in electron energy $\epsilon$
(Eq. \ref{powerlawel}). A specific model is defined by a unique functional
relation between the spectral index $\delta$ and the spectrum normalization
$\Phi_{\epsilon_0}$. The relation depends on some model parameters, which
are assumed to be constant during a flare or a subpeak. This enables us to
compare the model function with the observed dataset
$(\delta(t),\Phi_{\epsilon_0}(t))$.
The first scenario is based on \emph{ad hoc} assumptions, while
the second is derived from a stochastic electron acceleration model
considered e.g. by Benz (\cite{benz77}). Newer models along the
stochastic acceleration line, like the transit-time damping
model proposed by Miller at al. (\cite{miller96}), do not imply
a simple functional relationship, and therefore cannot be included at this
stage. The spectral evolution of transit-time acceleration must be
calculated numerically and is not yet available.
Petrosian \& Liu (\cite{petrosian04}) also give plots of electron spectra
calculated from their model of stochastic acceleration by parallel waves,
but do not predict the time evolution of the spectra either.
For a quantitative comparison between observations and theory, we
are limited to models making concrete and usable predictions on
the relation between spectral index and power-law normalization.

\subsection{The constant productivity scenario}

A class of models may be defined assuming that the \emph{productivity}
of the accelerator is constant above a threshold energy $\epstar$.
Either the electron acceleration rate $N$ (electrons s$^{-1}$) to
energies above $\epstar$ or the total power $P$ input to electrons
above $\epstar$ are held constant, but the acceleration process
evolves in such a way that flux and index vary. Although such processes
have never been proposed, they may fit some of the data.
Upon integrating Eq. (\ref{powerlawel}), the models give the following
relations between the electron spectral index $\delta$ and the electron
spectrum normalization
$\Phi_{\eps_0}$:
\begin{eqnarray}
\label{fixedn}
\Phi_{\eps_0}&=& \frac{N (\delta-1)}{\epstar}
   \left(\frac{\epstar}{\eps_0}\right)^{\delta}\quad \mbox{(constant rate model)} \\
\label{fixede}
\Phi_{\eps_0}&=& \frac{P (\delta-2)}{\epstar^2}
   \left(\frac{\epstar}{\eps_0}\right)^{\delta}\quad \mbox{(constant power model)}
\end{eqnarray}
The total power in Eq. (\ref{fixede}) is expressed in the somewhat unusual
unit of keV s~$^{-1}$. The power in erg s~$^{-1}$ can be easily obtained
multiplying $P$ by the conversion factor $1.602\cdot 10^{-9}$~erg~kev~$^{-1}$.

\subsection{The stochastic acceleration scenario}

The index-flux relation in stochastic acceleration was explored by Benz
(\cite{benz77}) and further elaborated by Brown \& Loran (\cite{brown85}).
In this model plasma waves accelerate
stochastically the electrons in a plasma slab.
From the diffusion equation for the electron
distribution function Benz gets the approximate relation (corresponding to
Eq. 20 in Benz, \cite{benz77}, with the spectral index transformed from the
electron \emph{number} distribution into the corresponding value for the
electron \emph{flux} distribution)
\begin{equation}
\label{deltabasicbenz}
\delta=-\frac{1}{2}+ \frac{8e^2\,\ln\Lambda n}{\pi W_0}+ \frac{\sqrt{2}
\epstar}{\pi e \sqrt{W_0 L}}=:-\frac{1}{2}+\mathbf{d}+\mathbf{e}\,,
\end{equation}
where $e$ is the electron charge, $\ln\Lambda$ the Coulomb logarithm,
$n$ the ambient plasma electron density, $L$ the length of the plasma
sheet and $W_0$ the { time-dependant spectral} energy density in the waves
causing the acceleration.
Following Brown \& Loran (\cite{brown85}), we recognize that the term
$\mathbf{d}$ can be written as $\mathbf{d}=\alpha \mathbf{e}^2$ with
\begin{equation}
\alpha=\frac{4 \pi e^4 \ln\Lambda n L}{\epstar^2}=\frac{L}{L_\mathrm{MFP}},
\end{equation}
where $L_\mathrm{MFP}$ is the Coulomb collisional mean free path of the
electrons in the plasma sheet. The acceleration process needs
$L_\mathrm{MFP}\gg L$ to be effective, and therefore $\alpha \ll 1$.
Benz argues that the total electron flux
$\Phi_\mathrm{TOT}$ above $\epstar$ is proportional to $W_0 L$, that is
\begin{equation}
\label{fluxpropwl}
\Phi_\mathrm{TOT}=KW_0L .
\end{equation}
Using Eqs. (\ref{deltabasicbenz}) and (\ref{fluxpropwl}) we get
the following relation $\Phi_{\epsilon_0}\leftrightarrow\delta$
\begin{equation}
\label{stocaccalpha}
\Phi_{\eps_0}=\left(\frac{\epstar}{\eps_0}\right)^{\delta}
   \frac{C(\delta-1)}{\epstar}\left(
   \frac{1+\sqrt{1+4\alpha\left(\delta+\frac{1}{2}\right)}}
   {2\left(\delta+\frac{1}{2}\right)}\right)^2,
\end{equation}
where $C=2K \epstar^2 / (\pi^2 e^2)$.
In the collisionless case $\alpha=0$ it simplifies to
\begin{equation}
\label{stocacc}
\Phi_{\eps_0}=\left(\frac{\epstar}{\eps_0}\right)^{\delta}
   \frac{C}{\epstar}\frac{\delta-1}
   {\left(\delta+\frac{1}{2}\right)^2}.
\quad\mbox{(stochastic acceleration model)}
\end{equation}

Figure \ref{model_spectra} shows the spectra for different values of
$\delta$ and their $\Phi_{\epsilon_0}(\delta)$ according to Eqs.
(\ref{fixedn}), (\ref{fixede}) and (\ref{stocacc}).
Interestingly, the model described by Eq. (\ref{stocacc}) has the special
property that all the spectra cross each other in a very narrow region of
the plot, therefore exhibiting a behavior very similar to the one given by the
\emph{pivot point model} described in the following.

The pivot point model assumes that all the non-thermal
power-law spectra in the time series cross each other in an unique
point. This pivot point is characterized by its energy $\epstar$ and its
flux $\phistar$. The following relation holds between the spectral 
index $\delta$ and the power-law normalization $\Phi_{\eps_0}$ at energy
$\eps_0$
\begin{equation}
\label{pivpoint}
\Phi_{\eps_0}=\phistar\left(\frac{\epstar}{\eps_0}\right)^{\delta}
\quad\mbox{(pivot point model)}
\end{equation}

The relation between $\delta$ and $\Phi_{\epsilon_0}$ is given by Eqs.
(\ref{fixedn}), (\ref{fixede}), (\ref{stocaccalpha}), (\ref{pivpoint})
for, respectively, the constant rate model, the constant power model,
the stochastic acceleration model and the pivot point model.
All models depend on two free parameters assumed to be constant during
a flare or an emission peak.
The threshold energy for the first 3 models and the energy of the
pivot point in the last model have been all represented by the same
symbol $\epstar$, such that it is easier to compare the different equations.
When more distinction is needed we will refer to them as
$\epstar^\mathrm{RATE}$, $\epstar^\mathrm{POWER}$, $\epstar^\mathrm{STOC}$
and $\epstar^\mathrm{PIV}$ for the different models.
The second parameter, denoted $N$, $P$, $C$, $\phistar$, respectively,
characterizes the flux normalization in the 4 models.

\subsection{Relations with the pivot point model}

We now compute the energy of the approximate pivot point
$\epstar^\mathrm{PIV}$ which results from a stochastic acceleration
model given by the parameters $(C,\epstar^\mathrm{STOC})$ with $\alpha=0$.
To compute the position of the pivot point, we find first the
position of the intersection point of two spectra given by
$\delta_1, \Phi_1=\Phi_{\epsilon_0}(\delta_1)$ and 
$\delta_2, \Phi_2=\Phi_{\epsilon_0}(\delta_2)$
in the stochastic acceleration model.
Since the spectra are straight lines in logarithmic representation,
using Eqs. (\ref{powerlawel}) and (\ref{stocacc}) it is straightforward
to find for the intersection $\eps_\mathrm{INT}$
\begin{equation}
\ln\frac{\eps_\mathrm{INT}}{\epstar^\mathrm{STOC}}
=\frac{\ln\Phi_2-\ln\Phi_1}{\delta_2-\delta_1}=
\frac{\ln\frac{\delta_2-1}{\delta_1-1}
     -2\ln\frac{\delta_2+\frac{1}{2}}{\delta_1+\frac{1}{2}}}
{\delta_2-\delta_1}
\end{equation}
To find the approximate pivot point, we take the limit 
of the previous expression for $\delta_2\rightarrow\delta_1$.
Putting $\delta_1=\delta$ and $\delta_2=\delta+\Delta$ with the
condition that $\Delta / \delta \ll 1$, we get to the first order
in $\Delta / \delta$
\begin{equation}
\ln\frac{\eps_\mathrm{INT}}{\epstar^\mathrm{STOC}}=
-\frac{\delta-\frac{5}{2}}{(\delta-1)(\delta+\frac{1}{2})}.
\end{equation}
Therefore two spectra with spectral index around $\delta$ will
have a common point at
\begin{equation}
\eps_\mathrm{INT}=\epstar^\mathrm{STOC}\cdot
\exp\left(-\frac{\delta-\frac{5}{2}}{(\delta-1)(\delta+\frac{1}{2})}\right)
=:\epstar^\mathrm{STOC}\cdot f(\delta).
\end{equation}
The function $f(\delta)$ depends weakly on delta for $\delta > 3$,
and therefore all the spectra will cross each other in a narrow
energy and flux range. The energy of the corresponding pivot point
is approximately
$\epstar^\mathrm{PIV}=\epstar^\mathrm{INT}\simeq\epstar^\mathrm{STOC}\cdot
f(6)=0.90\,\epstar^\mathrm{STOC}$.
Similar relations holds approximatively also for the other models:
\begin{eqnarray}
\eps_\mathrm{PIV}&\simeq&\epstar^\mathrm{RATE}\cdot
\exp\left(\frac{1}{\delta-1}\right)\\
\eps_\mathrm{PIV}&\simeq&\epstar^\mathrm{POWER}\cdot
\exp\left(\frac{1}{\delta-2}\right)
\end{eqnarray}

In the constant productivity models, the pivot point has an energy
which is slightly larger than $\epsilon_*$, and in the stochastic
acceleration model the pivot point energy is slightly lower than
$\epsilon_*$. Therefore in these models most accelerated electrons
have energies comparable with the pivot point energy.
On the other hand, the purely phenomenological pivot point model does not
request the presence of electrons at energies close to the pivot point
energy: the pivot point may be \emph{virtual} in the sense that the electron
energy distribution may be a power-law at higher energies and turnover at
an energy $\epsilon^\mathrm{TURN}>\epsilon^\mathrm{PIV}$.

\begin{table*}
\centering
\caption{Results of model fitting to the
observed evolution of the spectral index and non-thermal flux in the rise
and decay phases of the non-thermal emission peaks.}
\begin{tabular}{lcccccclc}
\hline\hline
Model         & Equation for                              & Phase & Nr. of & Nr. of phases & Nr. of phases        & Fitted   & Centroid of the & Half width of    \\
              &  $\Phi_{\epsilon_0}\leftrightarrow\delta$ &       & phases & in fit region & in 3-$\sigma$ region & variable & distribution$^{\rm a}$   & the distribution$^{\rm b}$ \\
\hline
Constant      & (\ref{fixedn})   & Rise  & 70 & 64 & 52 &  $N$        & $5.70\cdot 10^{35}$ el. s$^{-1}$            & 15  \\
rate          &                  &       &    &    &    &  $\epstar$  & 13.9 keV                                    & 1.6 \\
              &                  & Decay & 71 & 68 & 56 &  $N$        & $2.05\cdot 10^{35}$  el. s$^{-1}$           & 8.4 \\
              &                  &       &    &    &    &  $\epstar$  & 18.6 keV                                    & 1.5 \\
\hline				 									            
Constant      & (\ref{fixede})   & Rise  & 70 & 64 & 51 &  $P$        & $1.18\cdot 10^{37}$ keV s$^{-1}$            & 8.3 \\
power         &                  &       &    &    &    &  $\epstar$  & 13.6 keV                                    & 1.6 \\
              &                  & Decay & 71 & 67 & 55 &  $P$        & $5.85\cdot 10^{36}$ keV s$^{-1}$            & 4.5 \\
              &                  &       &    &    &    &  $\epstar$  & 17.6 keV                                    & 1.4 \\
\hline				 									            
Stochastic    & (\ref{stocacc})  & Rise  & 70 & 64 & 55 &  $C$        & $8.09\cdot 10^{36}$ s$^{-1}$                & 16  \\
acceleration  &                  &       &    &    &    &  $\epstar$  & 18.1 keV                                    & 1.7 \\
              &                  & Decay & 71 & 69 & 56 &  $C$        & $2.23\cdot 10^{36}$ s$^{-1}$                & 7.8 \\
              &                  &       &    &    &    &  $\epstar$  & 24.2 keV                                    & 1.4 \\
\hline
Pivot         & (\ref{pivpoint}) & Rise  & 70 & 63 & 53 &  $\Phi_*$   & $9.39\cdot 10^{34}$ el. s$^{-1}$ keV$^{-1}$ & 18  \\
point         &                  &       &    &    &    &  $\epstar$  & 17.5 keV                                    & 1.6 \\
              &                  & Decay & 71 & 67 & 56 &  $\Phi_*$   & $1.28\cdot 10^{34}$ el. s$^{-1}$ keV$^{-1}$ & 11  \\
              &                  &       &    &    &    &  $\epstar$  & 22.3 keV                                    & 1.4 \\
\hline
\end{tabular}
\label{tablemod}
%
%
{\footnotesize
\flushleft
(a) centroid of the best fit 2-dimensional Gaussian distribution \\
(b) expressed as a multiplicative factor \\
}
\label{list}
\end{table*}

%
\section{Comparison with the data}\label{comparison}
%

\subsection{Fits in $\delta$--$\Phi_{\epsilon_0}$ plane}

We proceed now to compare the observed evolution of the spectra
with the models presented in the previous section.
The dataset described in Section \ref{data} consists of discrete
time series of the electron spectral index
$\delta(t_i)$ and of the power-law normalization $\Phi_{\eps_0}(t_i)$ at
energy $\eps_0$ at the times $t_1,t_2,\dots,t_n$ for the 24 different flares.
For each model described in Section \ref{acc_scenario} we have
a relation between $\Phi_{\epsilon_0}$ and $\delta$,
and the comparison of the data is done by a least-square fitting of the
inverse function $\delta(\Phi_{\eps_0},\mathbf{P})$ to the observed pairs
$(\Phi_{\eps_0}(t_i),\delta(t_i))$, where $\mathbf{P}$ is the model
parameter vector. Since the spectral index $\delta$ varies over a smaller
factor than the flux normalization $\Phi_{\epsilon_0}$,
it is better to use the latter as independent variable for the fitting, and
therefore we use the inverse function $\delta(\Phi_{\eps_0})$ instead of
$\Phi_{\eps_0}(\delta)$.
The choice of $\epsilon_0$ is arbitrary, but we settled for $\eps_0=60$~keV,
consistent with the observed range of electron energies.
Paper I remarked that the scatter of the data is smaller on average
for time series belonging to rise or decay phases of
the non-thermal emission peaks. A given rise or decay phase is defined
as a series of at least three consecutive points showing an increase
or a decrease of the flux. Therefore we independently fit the model parameters
to each rise and decay phase.
Figure \ref{datamodel} shows $(\Phi_{60},\delta)$ points consequent in time
measured for three flares. 
The constant rate and stochastic acceleration models corresponding to the
best fit to the rise and decay phases are both shown in all graphs.
In each fit the model parameters are constant in time.
The curves corresponding to the two different models are nearly identical
because Eqs. (\ref{fixedn}) and (\ref{stocacc}) are functionally similar.
The two other models (constant power and pivot point) yield
curves (not shown in Fig.  \ref{datamodel}) which also are much
alike the ones shown.

\begin{figure*}
\centering
\includegraphics{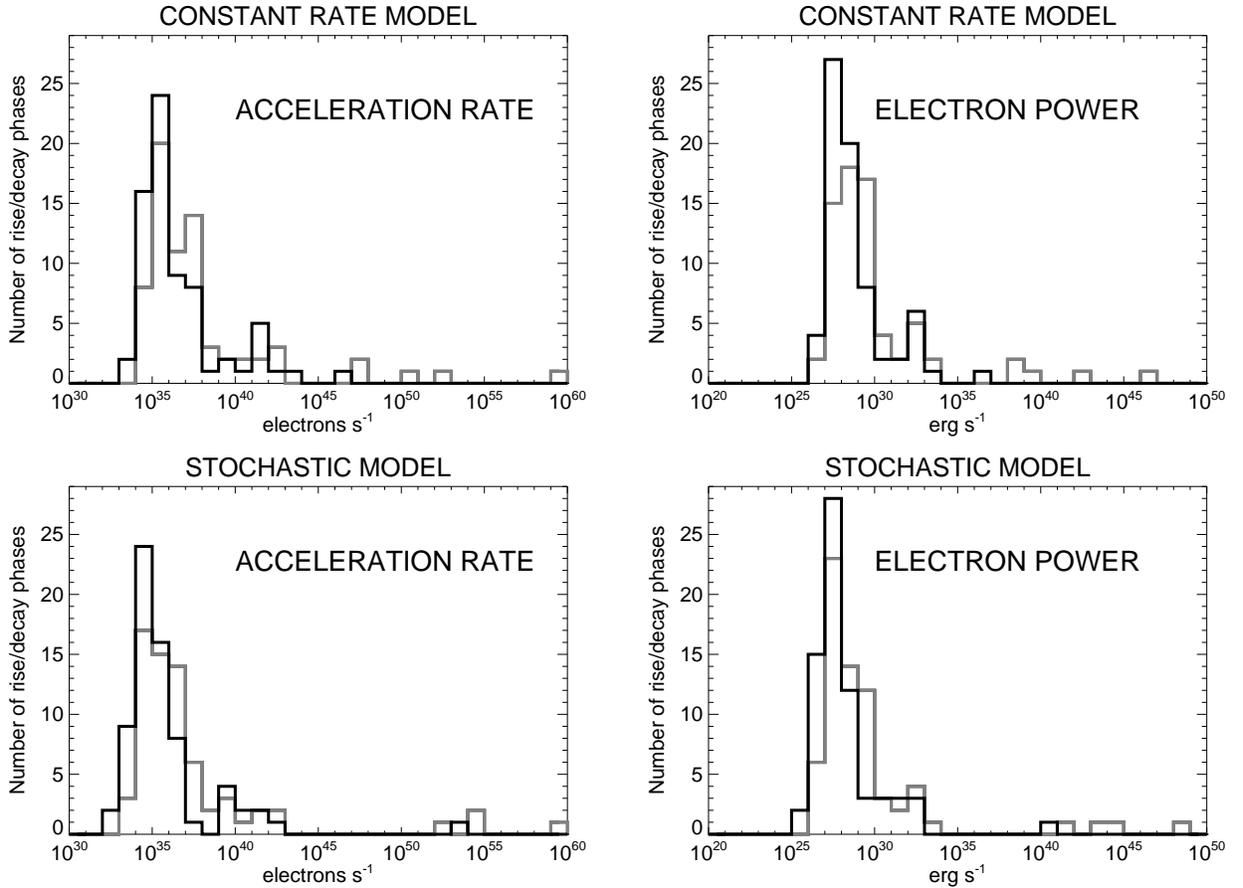}
\caption{Distribution of the electron acceleration rate (left column) and
the total power in the accelerated electrons (right column) for the constant
rate model (upper row) and the stochastic model (lower row). The distributions
are shown in gray for the rise phases and black for the decay phases.
The tail of the distribution to the right of the main peak contains events
with unphysically high values for the rates and powers.}
\label{ratepowerhisto}
\end{figure*}

\subsection{Comparing model parameters}

As a next step we compare the best-fit parameters between the different
models. Is one of the models preferable, yielding more plausible values?
Note that there is an evident correlation between the two parameters of each
model. It is due to the fact that the points $(\delta,\Phi_{60})$ of each
phase lie approximately on straight lines with different slopes, but all of
them passing near the point
$(\Phi_{60}=10^{31}\,\,\mbox{electrons s$^{-1}$ kev$^{-1}$},\delta=6.5)$.
This geometrical constraint in the $\delta$--$\Phi_{60}$ plane is the cause
of the correlation between the two model parameters: if $\epsilon_*$ is
small, $N$, $P$, $C$, or $\Phi_*$, respectively, must be large.
The distributions of the best-fit model parameters for the constant rate
and the stochastic acceleration model are presented in Fig. \ref{twodimdist}
separately for the rise and decay phases. Since the parameters vary
over several orders of magnitude, we chose a logarithmic representation.

In all the four models, we found that most of the parameters are concentrated
in a relatively narrow region, except for about 20--30\% of the points lying
far away from the peak of the distribution. The average value and standard
deviation of the distribution are poor estimators of the position and width
of the central peak in such a case. Instead, we characterized the
distributions by fitting two-dimensional Gaussian functions to the data.
This procedure yields a reasonable estimate for the position and width of the
distribution's peak. To further ensure that outliers do not have strong
influence, we restricted the fit to the region where $\eps_*>0.1$~keV and
the second fit parameter is smaller than $10^{45}$ (in each parameter's units).
The number of cases in the region to be fitted is given in Table
\ref{tablemod}, and its range is displayed in Fig. \ref{twodimdist}.
The model parameter distribution must be binned for the fitting procedure.
We chose bins centered on the median values of the two parameters and
with a width of 0.5 times the average deviation from the median.

In Fig. \ref{twodimdist} the contour lines corresponding to the
$1\sigma$, $3\sigma$ (thick line) and $5\sigma$ levels of the two-dimensional
Gaussian distribution are superimposed on the data.
We define the \emph{outliers} as being the points outside the $3\sigma$
contour line. The number of outliers is about 23~\% and does not vary
significantly in the different models and phases (Table \ref{tablemod}).

The outliers correspond to rise or decay phases where the spectral index
changes little as the flux increases or decreases. In the framework
of the pivot point model: if the pivot point for the spectrum lies 
at very low energy and high flux, it is possible to have large
variations at high energies with small changes in the
spectral index. The other 3 models behave in a fashion similar to the pivot
point model (as shown in Fig. \ref{model_spectra}), so this explication
also applies for them.

With the exception of the pivot point model, the parameters of the other
models have a direct physical meaning. We can check if they lie in an
acceptable range when compared with other observations and generally
accepted values on solar flares.
For the sake of simplicity and uniformity, we derive first the total power
injected in the accelerated electrons (in erg s$^{-1}$) and the
acceleration rate of electrons (in electrons s$^{-1}$) for the different
models. The following equations express the electron acceleration rate
$N=\int_{\epstar}^\infty\Phi(\epsilon)\,\mathrm{d}\epsilon$
as a function of the model parameters:
\begin{equation}
N^\mathrm{RATE} = N,
\end{equation}
\begin{equation}
N^\mathrm{POWER}= \frac{P(\delta-2)}{\epstar (\delta-1)},
\end{equation}
\begin{equation}
\label{nstoc}
N^\mathrm{STOC} = \frac{C}{(\delta+\frac{1}{2})^2}.
\end{equation}
The following equations express the power
$P=\kappa\int_{\epstar}^\infty\epsilon\,\Phi(\epsilon)\,\mathrm{d}\epsilon$
as a function of the model parameters: 
\begin{equation}
P^\mathrm{RATE} = \kappa\, N\epstar\frac{\delta-1}{\delta-2},
\end{equation}
\begin{equation}
P^\mathrm{POWER}= \kappa\, P, \\
\end{equation}
\begin{equation}
P^\mathrm{STOC} = \kappa\, C\epstar
   \frac{\delta-1}{(\delta-2)(\delta+\frac{1}{2})^2},
\end{equation}
where $\kappa=1.602\cdot 10^{-9}$~erg~kev~$^{-1}$.
Figure \ref{ratepowerhisto} shows the distribution of the acceleration rate
and the electron power for the constant rate model and the stochastic model,
for both the rise phases (gray) and the decay phases (black).
The distributions show a central peak and an extended tail to the right.
Rise and decay phases in the tail represent about the same subset
in all models, requiring extremely high values for both acceleration rate
and power. This is due to the fact that these phases
fit models having a very low threshold energy $\epstar$, and the integral of
the spectrum diverges in the limit $\epstar\rightarrow 0$. The low value of
$\epstar$ is necessary to account for the events where the spectral
index varies slowly during the phase.

What are the maximum values acceptable in reality?
The interpretation of hard X-ray observations of solar flares requires a
large number of accelerated electrons.
In his review on the flare mechanism, Sweet (\cite{sweet69}) already
requires $10^{36}$ electrons s$^{-1}$ to account for the observed hard X-ray
emission. Brown \& Melrose (\cite{brown77}) derive
a requirement of $5\cdot 10^{36}$ electrons s$^{-1}$ accelerated above 25 keV.
Miller (\cite{miller98}) cites $10^{36}$--$10^{37}$ electron s$^{-1}$ above
20 keV.
Recent RHESSI observations of X class solar flares yield similar
values (e.g. Holman et al. \cite{holman03}, Saint-Hilaire \& Benz
\cite{saint-hilaire05}). If the non-thermal electron
spectrum extends to energies lower than 20 keV, this number could
be higher. This usually does not contradict the observations, since during
a large flare, the photon flux at energies lower than 20 keV is typically
dominated by thermal emission.
From a theoretical point of view the number of flare electrons and energy
available in the active region environment is limited. 
Some mechanism for electron replenishment needs to be
operative but is often left unspecified.
Therefore it is not clear how the sustainable electron acceleration rate
is limited.
As an upper limit for comparison with our results, we take as the highest
reported electron acceleration rate $N^\mathrm{MAX}=10^{37}$ electron
s$^{-1}$. 
There is also a physical limit on the power that can be injected into the
accelerated electrons. As a generally acceptable maximum power
$P^\mathrm{MAX}$, we will assume $10^{29}$ erg s$^{-1}$ in agreement with
the above cited authors.

Some of the derived model parameters are highly implausible, requiring
acceleration rates and power input far exceeding the above limits.
In Fig. \ref{twodimdist} the limit for $\delta=6$ is indicated in the
$N$-$\epsilon_*$ and $C$-$\epsilon_*$ planes. Table \ref{tablemod} gives the
number of cases in total, as well as in Fig. \ref{twodimdist} and in the
region used for fitting a Gaussian distribution.
Figure \ref{twodimdist} indicates that most rise and decay phases below
the limit of plausibility are within the 3-$\sigma$ limit of the distribution,
and thus corroborate the Gaussian fit. Its centroid and half-widths
(in log presentation) are given in Table \ref{tablemod}.

{ Since we follow the time evolution of the spectra from the onset of the
non-thermal emission, there is a large difference of more than 2 order of
magnitude between the value of the minimum and maximum normalization
$\Phi_{60}$.
Surprisingly, the clustering of phases in the $N$-$\epsilon_*$ and
$C$-$\epsilon_*$ planes (Fig. \ref{twodimdist}) indicate that most flare
peaks can be interpreted by models within a relatively small range of
threshold energies $\epsilon_*$.
The second model parameters $N$, $P$, $C$, $\phistar$ are spread
upon a larger range. They include the effect of the different sizes of the
non-thermal emission peaks of our sample.}
Note that $\epsilon_*$ is smaller in the rise phase than in the decay of
a flare peak on average. On the other hand, the normalization is larger in
the rise phase on average.

{ Gan (\cite{gan99}) reported the presence of a pivot point in
the SMM/GRS spectra of two X-class flares. His results of 41 keV and
77 keV for the energy of the pivot point of the electron spectra are
larger than ours. This may be due to the fact that he analyzed
larger events, or that he fitted energies mostly above our energy range,
thus possibly above an high-energy break in the spectrum.}

The value of the { spectral} wave energy density $W_0$ in the stochastic
acceleration model can be computed.
From Eqs. (\ref{fluxpropwl}), (\ref{nstoc}) and $C=2K \epstar^2 / (\pi^2 e^2)$
we get:
\begin{equation}
W_0L=\frac{2\epstar^2}{\pi^2e^2(\delta+\frac{1}{2})^2}
\end{equation}
Using the average values of $\epstar=20$~keV and $\delta=6.4$, this amounts to
\begin{equation}
W_0\simeq 2\times 10^{-6}\left(\frac{10^7 \mathrm{cm}}{L}\right)
\,\,\mathrm{erg}\,\,\mathrm{cm}^{-2}\,.
\end{equation}
This value is somewhat lower than the ones cited in Benz (\cite{benz77}) and
Brown \& Loran (\cite{brown85}), because we have used a lower value of
$\epstar$ and
a larger value of $\delta$. In the stochastic model, the total energy density
$U_\mathrm{T}=\int W(k)\,\mathrm{d}k$ in the
turbulent waves can be computed by
\begin{equation}
U_\mathrm{T}\simeq \frac{W_0}{\lambda_\mathrm{D}}
\simeq 3\times 10^{-5}\sqrt{\frac{n_e}{10^{10}\,\mathrm{cm}^{-3}}}
\,\mathrm{erg}\,\mathrm{cm}^{-3},
\end{equation}
where $\lambda_\mathrm{D}$ is the Debye length.
It is well below the magnetic energy density, which is larger than unity.

%
\section{Discussion}
%

The four models presented in Section \ref{acc_scenario} are based on different
assumptions, but have similar functional relationships between the electron
spectral index $\delta$ and the non-thermal spectrum normalization
$\Phi_{\eps_0}${, as required by observations}.
As a consequence, it is possible to visualize the spectral
evolution of all these models by the presence of a common \emph{pivot point}
in the non-thermal electron spectra during the rise or decay phase of each
hard X-ray peak. This general property however, inhibits selecting the best
model from spectral fits.

The comparison of the observed
evolution of the non-thermal spectrum during rise and decay phases of a
non-thermal emission peak with the models yields best-fit model parameters.
Considering the simplicity of the assumed models it is surprising that
the distribution of the model parameters is reasonable for about 77\%
of the 141 observed rise/decay phases, but yields an unphysically high electron
acceleration rate for the rest of the observed events. The numbers of outliers
is not statistically different between the models.

The events in the
unphysical region of parameter space manifest themselves as the enhanced
tail of the two-dimensional distribution shown in Fig. \ref{twodimdist}.
The simple models do not succeed in describing these events, which all
exhibit a slowly varying spectral index.
These models would totally fail to reproduce events with constant spectral
index.
The mathematical reason of the failure is that these model just assume
power-law behavior of the observed electron distribution above a fixed
threshold energy $\epstar$.
To account for the events with slowly varying spectral index, the fit
converges toward very low values of $\epstar$, whereas the models can be
justified only in the case where $\epstar$ is larger than about a few keVs.
More sophisticated models, like the one proposed by Miller et al.
(\cite{miller96}) or Petrosian \& Liu (\cite{petrosian04}), explicitely
accelerate non-thermal electrons \emph{out of a thermal distribution}.
They do not show a power-law behavior at low energy and therefore
should be less susceptible to such low-energy divergencies.

The best-fit model parameters show an asymmetry between the rise and
decay phases of the emission peaks. Such an asymmetry was already reported
in Paper I. In the constant productivity scenario, this would correspond to a
reduced productivity of the accelerator in the decay phase.
Similarly, the fitting to the stochastic acceleration model suggests a
lower electron acceleration rate on average for the decay phase.

%
\section{Conclusion}
%

We compared observations (described in Paper I) from RHESSI on the evolution of
the normalization and spectral index of the non-thermal component during
emission peaks of solar flares with model predictions by means of fittings
the model parameters to the observed values for the electron spectral index
$\delta$ and the spectrum normalization $\Phi_{60}$ at 60~keV (as shown in
Fig. \ref{datamodel}).
All the models selected for this study, described in Section
\ref{acc_scenario}, feature the soft-hard-soft behavior for a single flare
peak and fit well the observed spectral evolution of flare peaks.
We have shown in Section \ref{acc_scenario} that they have a
functional dependence between $\delta$ and $\Phi_{60}$ which closely resembles
the one given by the pivot point model (Eq. \ref{pivpoint}).
For this reason the data cannot discriminate between models.

While it is possible to fit reasonably well the models to the observed data,
the resulting model parameters imply unphysically high electron acceleration
rates and energies for the 20--30 \% events were the
spectral index changes slowly as the flux rises, thus showing less
prominent soft-hard-soft behavior (Fig. \ref{ratepowerhisto}).
This is due to the fact that the models need a very low threshold energy
$\epstar$ (that is, the energy above which electrons are accelerated,
corresponding approximately to the energy of the pivot point) to account for
low rates of change of the spectral index.
A low value for $\epstar$ enables the model to provide reasonably good fits
to the data, but will often lie outside the range of validity of the models
studied. For instance, the stochastic acceleration model will break down
for electron energies comparable to the thermal energy of the medium in
which the electron are accelerated, since the interaction of the accelerating
waves with the thermal component was not considered for this simple model.
The assumption of a fixed $\epstar$ during a peak may also be challenged.
Allowing $\epstar$ to change during the observed events could reduce the
excess acceleration rate values.

The { quantitative} relation between the observed spectral index and the
normalization of the spectrum exploited here has the advantage
that it does not depend explicitely on the time evolution of the two variables.
Therefore it is comparatively easier to compare with model predictions
than the full-fledged time evolution of the spectrum.
The $\delta\leftrightarrow\Phi$ relation puts an additional requirement on
the acceleration mechanism.

The comparison between observations and models proposed
here can be done with real data in a straightforward way, and
it does provide { interesting} results for relatively simple
theoretical models: a majority of flare peaks (about 77~\% of the rise and
decay phases) can be well fitted by these simple models within a compact region
of parameter space.
Nevertheless, the rest (outliers with unphysical model parameters) point to
the fact that more degrees of freedom are necessary for interpretation.
While there is a wealth of data available thanks to the RHESSI mission,
we lack concrete predictions by more complex acceleration models on the
behavior of the spectral index during flares.

\begin{acknowledgements}
The analysis of RHESSI data at ETH Zurich is partially supported by the Swiss
National Science Foundation (grant nr. 20-67995.02). This work relied on the
RHESSI Experimental Data Center (HEDC) supported by ETH Zurich
(grant TH-W1/99-2). We thank the many people who have contributed to the
successful operation of RHESSI and acknowledge P. Saint-Hilaire, G. Emslie,
K. Arzner for helpful discussions.
\end{acknowledgements}

%

%
\end{document}